\documentclass{emulateapj}
\usepackage{graphicx,epsfig,psfig}
\usepackage[normalem]{ulem}
\usepackage[dvipsnames]{xcolor}
\usepackage[]{inputenc,amssymb}
\usepackage{amsmath}
\usepackage{color}
\usepackage{natbib}

\definecolor{webgreen}{rgb}{0,.5,0}
\definecolor{webbrown}{rgb}{.6,0,0}
\usepackage[pdfpagelabels]{hyperref}
\hypersetup{%
   colorlinks=true,hyperfootnotes=false,%
   breaklinks=true,%
   plainpages=false, bookmarksnumbered, bookmarksopen=true,
   bookmarksopenlevel=1,%
   urlcolor=webbrown, linkcolor=RoyalBlue, citecolor=webgreen,
   }
\usepackage[toc,page]{appendix}

\defcitealias{sha12}{S12}

\newcommand{\myemail}{deovrat@physics.iisc.ernet.in}

\newcommand{\tr}{$t_{\rm cool}/t_{\rm ff}$}

\newcommand{\be}{\begin{equation}}
\newcommand{\ee}{\end{equation}}
\newcommand{\ba}{\begin{eqnarray}}
\newcommand{\ea}{\end{eqnarray}}

\def \tr {$t_{\rm cool}/t_{\rm ff}$}

\shorttitle{BCG in Galaxy Clusters}
\shortauthors{D. Prasad, P. Sharma, A. Babul}

\begin{document}

\title{Cool-Core Clusters : Role of BCG, Star Formation \& AGN-driven turbulence}

\author{Deovrat Prasad$^1$}\email{\myemail} \author{Prateek Sharma$^1$}\email{prateek@physics.iisc.ernet.in}
\affil{$^1$Joint Astronomy Program and Department of Physics, Indian Institute of Science,
    Bangalore, India 560012}
\author{Arif Babul$^2$}\email{babul@uvic.ca}
\affil{$^2$Department of Physics and Astronomy, University of Victoria, Victoria, BC V8P 1A1, Canada}

\begin{abstract}
Recent analysis shows that it is important to explicitly include the gravitational potential of the central brightest central galaxy (BCG) to infer 
the acceleration due to gravity ($g$) and the free-fall time ($t_{\rm ff} \equiv [2r/g]^{1/2}$) in cool cluster cores. Accurately measuring $t_{\rm ff}$ is
crucial because according to numerical simulations 
cold gas condensation and strong feedback occur in cluster cores with min($t_{\rm cool}/t_{\rm ff}$) below 
a threshold value close to 10. Recent observations which include the BCG gravity show that the observed threshold in min($t_{\rm cool}/t_{\rm ff}$) lies
at a somewhat higher value, close to 10-30; there are only a few clusters in which this ratio falls much below 10.
 In this paper we compare numerical simulations of  feedback AGN (Active Galactic Nuclei)  jets interacting with the intracluster medium (ICM),
with and without a BCG potential. We find that, for a fixed feedback efficiency, the presence of a BCG does not significantly affect the temperature 
but increases (decreases) the core density (entropy) on average. 
Most importantly, min($t_{\rm cool}/t_{\rm ff}$) is only affected slightly by the inclusion of the BCG gravity. 
Also notable is that the lowest value of min($t_{\rm cool}/t_{\rm ff}$)
in the NFW+BCG runs are about twice larger than in the NFW runs because of a shorter time for feedback heating (which scales with the free-fall time) in the former.
We also look at the role of depletion of cold gas due to star formation and  show that 
it only affects the rotationally dominant component (torus), while the radially dominant component (which regulates the 
feedback cycle) remains largely
unaffected. Stellar gas depletion also increases the duty cycle of AGN jets. 
The distribution of metals due to AGN jets in our simulations is predominantly along the jet direction and the radial spread of metals is less compared to the observations.
We also show that the turbulence in cool core clusters is weak, consistent with recent {\it Hitomi} results on Perseus cluster.
\end{abstract}


\section{Introduction}
Dense X-ray emitting plasma in cool cluster cores
 is  susceptible to thermal fragmentation,
leading to the formation of a multiphase medium consisting of cold dense clouds condensing from the hot ICM.
The infall and accretion of these cold clouds onto the central super massive black hole (SMBH) powers the AGN outbursts that maintain the ICM (\citealt{piz05}) in rough thermal balance. Early idealized simulations predicted that cold gas stochastically condenses
out of the hot ICM if the minimum in the ratio of the cooling time to the free-fall time (min$[t_{\rm cool}/t_{\rm ff}]$) falls below a threshold 
close to 10 (\citealt{mcc12, sha12}). This cold gas is expected to lose angular momentum due to cloud-cloud collisions and due to drag imparted by the hot 
gas, fall inwards and fuel AGN outbursts (\citealt{gas13,dev17}). The
feedback process is self-regulatory with phases dominated by radiative cooling and jet heating
(\citealt{mcn05, raf06}).
 Several recent feedback jet simulations evolved over cosmological timescales are now able to reproduce the gross observed properties of cool cluster cores
 (\citealt{gas12b, li15, dev15,yan16}).

Since AGN feedback  is triggered by the precipitation of cold gas from the hot ICM (\citealt{sha12, voi15b}), the
feedback process is sensitively dependent on the properties of X-ray emitting gas in cluster cores. Recent works like \citet{voi15,hog17,hog17a} highlight the importance ofexplicitly including the central brightest cluster galaxy (BCG) to determine  the acceleration due to gravity ($g$) and free-fall time ($t_{\rm ff} \equiv [2r/g]^{1/2}$) in cluster cores. Most cool core clusters have BCGs at their centers whose gravity dominates the gravity due to the dark matter halo within the central 20-30 kpc. \citet{hog17a} argue that including the BCG gravity increases min(\tr) in most cool core clusters above 10, and hence observations are in tension
with the simulation results that find min(\tr) drop down to a few (albeit for a short time). In this paper we test if including the BCG potential changes the value of min(\tr) in the jet-ICM simulations, 
and compare our simulation results with observations.  

The AGN-ICM coupling can happen through shocks (\citealt{fab03, li16}), turbulence
(\citealt{zhu14}), mixing (\citealt{ban14,hil16}), entrainment (\citealt{mcn05,pop10}), cosmic rays and thermal conduction (\citealt{voig04,guo08,sha09}) or a combination of these processes (\citealt{cie18}). However, the relative importance of these various processes in heating the cluster core is not clear. 
Among these mechanisms, turbulent heating seems to be ruled out by recent {\sl Hitomi} observations of the Perseus cluster (\citealt{hit16}), which show that the turbulence level in the cluster core is weak. 
Even if turbulent heating may not be the dominant mechanism for core heating,  turbulent mixing and diffusion still plays an important role in core thermodynamics and in transporting out the freshly created metals in star-forming cool cluster cores. Consequently another aim of our paper is to compare AGN-driven turbulence in our simulations with the observational constrains on ICM turbulence.

In addition to maintaining rough thermal equilibrium in cool core clusters, AGN jets, as they rise buoyantly to 100s of kpc, also play a potential role in distributing metals 
by entraining metal enriched gas (\citealt{pop10, rev08}) from star forming inner regions of the cool-core clusters. 
Observations show that in nearby cool core clusters the central $\sim 100$ kpc have a sharply rising metallicity 
while the outer regions have a constant metallicity (\citealt{tam04,fuj08,sim11,wer13}). On the other hand, in non-cool clusters the 
metallicity in the cluster core increases only marginally with a decreasing radius, with the outer regions having a fixed metallicity similar to cool-core 
clusters (\citealt{lec08}). Further, the metallicity within the central 100-150 kpc evolves with redshift ($z<1.4$) in cool core clusters 
while the outer regions do not show any evolution (\citealt{ett15}). 

Observations of several cool core clusters show that the metallicity is high beyond the cluster core along the AGN jet direction (\citealt{sim09, kir09, osu11}) 
as compared to the perpendicular direction.
This suggests that the 
cavities created by AGN jets are able to carry the metal enriched gas to regions beyond the star forming cluster cores. Further, as the AGN outflow is bi-conical in nature,
the isotropic distribution of  metals (at a rather large value $\approx 0.3 Z_\odot$) 
in cluster outskirts suggests that the metal enrichment of ICM out to the virial radius happened at early times, by when most of the 
metals in the universe were already produced.  We include a crude model for metal injection 
by injecting metallicity within small biconical jet source regions rather than following the stellar distribution in the BCG. Metallicity at $\sim$100 kpc should be affected 
mainly by transport due to AGN jets rather than by injection. Moreover, we only consider metal production in the cluster center and ignore the dominant process of early 
enrichment in which metals were produced far from the BCG within the galaxies that merged to form the eventual cluster. The aim of our metallicity study is limited -- 
to quantify the dispersal of metals only due to AGN jets in realistic cool cores. 

This paper is organized as follows. In section \ref{sec:NumMet} we present the numerical setup followed by analysis methods. Section 
\ref{sec:res} presents the key results of our 3-D simulations. We compare the simulations with and without the central BCG in detail. 
We  provide a quantitative comparison between the BCG+NFW runs with and without stellar depletion. We also analyze hot gas 
velocity distribution and turbulence in cluster cores, and discuss the nature of metal distribution due to AGN jets.  In 
section \ref{sec:dis} we compare our results with observations and discuss their implications. We conclude with a brief summary in section 
\ref{sec:con}.

\section{Numerical Setup}
\label{sec:NumMet}
We modified the PLUTO MHD code (\citealt{mig07}) to simulate AGN feedback in galaxy clusters. We solve the standard hydrodynamic equations in
spherical coordinates ($r$, $\theta$, $\phi$) with cooling, external gravity and mass and momentum source terms due to AGN jet feedback, as
described in section 2 of \cite{dev15}. 
We explore the effects of BCG at the centre of galaxy cluster by including the BCG potential along with the usual NFW potential.
We use the same feedback prescription as \cite{dev15}, with a fixed feedback efficiency, $\epsilon=\dot{M}_j v_j^2/\dot{M}_{\rm acc}c^2 =5\times10^{-4}$ (see Eq. 6 in \citealt{dev15}), where $\dot{M}_j$ is the jet mass loading rate, $v_j~(=0.1 c$, $c$ is the speed of light) is the injected jet velocity, and $\dot{M}_{\rm acc}$ is the accretion rate through the radial inner boundary of the domain at $r_{\rm in}=0.5$ kpc. These parameters are somewhat different from our previous papers (\citealt{dev15,dev17}) but our results are qualitatively similar.

\subsection{Gravitational Potential}
\label{sec:gpt}
The dark matter halo mass ($M_{200}$) for all our runs is $7 \times 10^{14} M_\odot$.
One of our runs uses only the NFW gravitational potential (\citealt{nav97}). For the other two runs the external gravitational potential is the sum of two
different potentials: 1) NFW
dark matter potential, and 2) a singular isothermal sphere (SIS) for the central brightest cluster galaxy,
\begin{equation}
\Phi = \Phi_{\rm NFW} + \Phi_{\rm SIS}.
\end{equation}

The singular isothermal sphere potential (SIS) for the brightest cluster galaxy is given by:
\begin{equation}
\Phi_{\rm SIS} (r) = 4 \pi G \rho_0 a_0^2  \ln(r/a_0),
\end{equation}
 where $\rho_0 = 1.67\times 10^{-23}~{\rm g} {\rm cm}^{-3}$ and $a_0=3$ kpc. The isothermal sphere circular velocity $V_c= \sqrt{4\pi G \rho_0 a_0^2} = 350$
 km s$^{-1}$. The circular velocity is in the range of $V_c$ observed in the cluster sample of \cite{hog17} (note that this paper uses the equivalent stellar 
 velocity dispersion $\sigma_\star=V_c/\sqrt{2}$).

\subsection{Grid, Initial and Boundary Conditions}
\label{sec:bc}
We perform our simulations in spherical coordinates with $0 \leq \theta \leq \pi$, $0 \leq \phi \leq 2\pi$, and $r_{\rm min} \leq r \leq r_{\rm max}$, with $r_{\rm min} = 0.5$ kpc  and $r_{\rm max} = 500$ kpc. We use a logarithmically spaced grid in radius, and an equally spaced grid in $\theta$ and $\phi$.

The outer electron number density is fixed to be $n_e =7\times 10^{-4}$ cm$^{-3}$ . Given the entropy profile with a core (Eq. 7 in \citealt{dev15}) and the outer density, we solve for hydrostatic
equilibrium and obtain the density and pressure profiles in the gravitational potential (for details see \citealt{dev15}). As in \cite{dev15}, we introduce small (maximum over-density is 0.3) isobaric
density perturbations on top of the smooth density. The gas is allowed to cool to 50 K unlike \cite{dev15} where cooling was cut-off at $10^4$ K.

We apply outflow boundary conditions at the inner radial boundary, where gas is allowed to go out of the computational domain but not allowed to enter it. We fix the density and pressure at
the outer radial boundary to the initial value and the gas is not allowed to flow in/out of the outer radial boundary.
Reflective boundary conditions are applied in $\theta$ (with the sign of $v_{\phi}$ flipped at
the poles) and periodic boundary conditions are used in $\phi$.

\subsection{Stellar Gas Depletion}
\label{sec:stell}
One of our runs (see Table \ref{tab:tab1}) implements a crude model for mass depletion of cold gas due to star formation.
To simulate the removal of cold gas due to star formation, we deplete the cold gas with temperature $T < 0.005$ keV ($\approx 5\times10^4$ K) and density
$\rho > 10^{-24}$ g cm$^{-3}$ using a sink term in the mass conservation equation
\begin{equation}
\label{eq:mass_source}
\frac{\partial \rho}{\partial t}  + \nabla .(\rho {\bf v}) = S_\rho - D_{\rho}, 
\end{equation}
where the depletion term $D_\rho = \rho/\tau$ 
and $\tau=200$ Myr is the gas depletion timescale (this is on the lower side of the range seen in observations; e.g., see Fig. 9 of \citealt{pul18}).
Note that there is a large uncertainty in the determination of the star formation rate and hence $\tau$ (e.g., see \citealt{mit15}). Our choice of $\tau$
is such that our cold gas mass is in a range consistent with observations.  
Here, $S_\rho$ is the usual AGN jet mass source term as in
 Eq. 1 of \cite{dev15}. Note that, unlike here, in \citet{dev17} we only accounted for stellar depletion in post processing.
 We do not consider feedback due to star formation because it is sub-dominant compared to AGN feedback in massive halos.

\subsection{Metallicity}
\label{sec:met}

Using our realistic cool core simulations, we also wish to study the jet-driven transport of metals produced recently (after majority of stars within the cluster 
have already formed and the cluster with a cool core is assembled) in the BCGs of cool-core clusters.
To quantify the spread of metals in the ICM, we evolve the passive scalar equation with a source term in the jet mass source region
\be
\label{eq:metals}
\frac{\partial Z}{\partial t} + {\bf v} \cdot \nabla Z = Z_j \frac{S_\rho}{\rho},
\ee
where $Z$ is the metallicity defined as the ratio of metal mass and total gas mass (normalized to the solar value, $Z_\odot$), 
$Z_j$ 
 is the normalization of the
jet metallicity and $S_\rho$ is the jet mass source term (see Eq. \ref{eq:mass_source}). The jet metallicity $Z_j$ is somewhat arbitrary as our focus is 
on the spatial spread of metals due to AGN jets rather than the actual value of metallicity. We choose $Z_j=100$ as it gives a reasonable metallicity values
for the simulated ICM. This value is also justified by considering the mass loading factor of AGN jets relative to the star formation rate in the BCG.
From our jet feedback prescription, $\dot{M}_j = (\epsilon c^2/v_j^2 )
\dot{M}_{\rm acc} = 0.05 \dot{M}_{\rm acc}$ for our parameters.\footnote{As emphasized in \citet{dev17}, $\dot{M}_{\rm acc}$ depends on the choice of $r_{\rm in}$.
Current cluster simulations simply do not have the resolution to directly simulate accretion on to the SMBH.} 
The mass accretion rate at 0.5 kpc is $\sim 5 M_\odot~{\rm yr}^{-1}$ (see Table \ref{tab:tab1}). 
On the other hand the average star formation rate (SFR) is expected to be a few 10s of $M_\odot$yr$^{-1}$. 
Since all the metals produced due to star formation in the BCG are deposited in the jet region in our simulations, 
$Z_j \sim ({\rm SFR }/\dot{M}_{\rm acc})\times (\dot{M}_{\rm acc}/\dot{M}_j) \approx 5 \times 20 =  100$ is a reasonable order 
of magnitude normalization for the metal source term. 

In our metallicity profiles (c.f. Figs. \ref{fig:ang} \& \ref{fig:mrm}) we add $0.3 Z_\odot$ to account for the ambient metallicity close to the viral radius due to
early enrichment. The radial spread of metals due to AGN jets 
is quantified using,
\begin{equation}
\label{eq:metal_radius}
 Z(r) =  \frac{\int_\theta\int_\phi Z(r,\theta,\phi)\rho \sin\theta d\theta d\phi}{\int_\theta\int_\phi \rho \sin\theta d\theta d\phi},
\end{equation}
where $r$ is the radius.  
Similarly, the angular distribution of metallicity due to AGN jets is quantified by
\begin{equation}
\label{eq:ang_dist}
Z(\theta) = \frac{\int_r\int_\phi Z(r,\theta,\phi)\rho r \sin\theta dr d\phi}{\int_r\int_\phi \rho r\sin\theta dr d\phi}.
\end{equation} 

\section{Results}
\label{sec:res}

 In this section we describe the important results of our simulations. Table \ref{tab:tab1} lists all our runs.  We show that the inclusion of BCG potential does not 
 affect the cluster temperature but affects the average electron number density, $t_{\rm cool}/t_{\rm ff}$ and entropy profiles 
 in the core. We study the effect of stellar cold gas depletion on cluster evolution.  We also show that the turbulence level in cluster cores is weak,
 consistent with the recent {\sl Hitomi} results. We find that the metal distribution is anisotropic and too narrow in radius as compared to the observations. 
\begin{table*}
\caption{List of runs}
\resizebox{1.0 \textwidth}{!}{%
\begin{tabular}{c c c c c c c c c}
\hline
Run & $r_{\rm in}$  & $r_{\rm out}$ & $\epsilon$ & run time & $\dot{M}_{\rm acc}$ & $M_{\rm cold}$ & $\tau$ & fraction of time for which \\
       & (kpc) 	       & (kpc) & & (Gyr) &  ($M_{\odot}$yr$^{-1}$)    &  ($10^{10} M_\odot$) &  (Myr) & min$(t_{\rm cool}/t_{\rm ff})<$10, 5 \\
\hline
NFW & 0.5 & 500 & $5\times10^{-4}$ & 4 & 4.4 & 2.0 & $\infty$& 47\%, 14\%  \\
NFW+BCG & 0.5 & 500 & $5 \times 10^{-4}$ & 4 & 6.7  & 4 & $\infty$& 55\%,14\%  \\ 
NFW+BCGd$^\dag$ & 0.5 & 500 & $5 \times 10^{-4}$ & 4 & 7.1  & 0.2 & 200 & 77\%,19\%  \\ 
\hline
\end{tabular}}
\label{tab:tab1}
\\
\begin{flushleft}
The $M_{200}$ for all the runs is $7\times10^{14}$ $M_\odot$. The resolution of all runs, done in spherical $(r_{\rm min} \leq r \leq r_{\rm max}, 0\leq \theta \leq \pi, 0 \leq \phi \leq 2\pi)$ coordinates,
is $256\times128\times32$. A logarithmic grid is used in the $r-$ direction, and a uniform one in others. $\dot{M}_{\rm acc}$ is the average (cold+hot)
mass accretion rate across $r_{\rm in}$ from 0 to 4 Gyr; $M_{\rm cold}$ is the total cold ($T<0.005$ keV) gas mass in the simulation domain by the end of the
run. \\
$^\dag$ `d' at the end of the label stands for depletion of cold gas.
\end{flushleft}
\end{table*}

\subsection{NFW vs NFW+BCG potential}

\begin{figure}
 \includegraphics[width=0.5\textwidth]{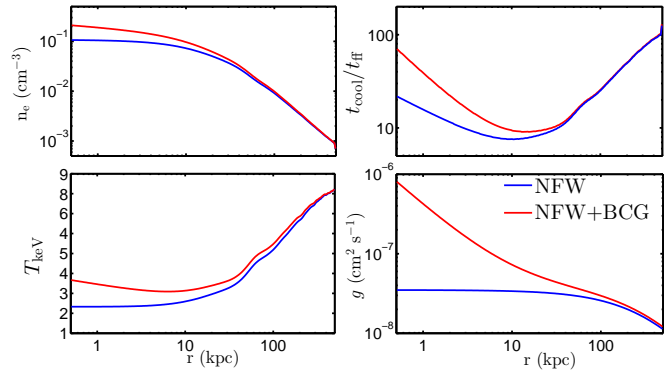}
  \caption{Comparison of the initial electron number density, temperature, $t_{\rm cool}/t_{\rm ff}$ and gravitational acceleration ($g$)
  profiles for the NFW (blue line) and NFW+BCG (red line) runs. The effect of the BCG is felt only within the central 50 kpc.}
  \label{fig:comp}
\end{figure}

\subsubsection{Average 1-D Profiles}

Figure \ref{fig:comp} shows the initial angle-averaged profiles of electron number density, temperature, $t_{\rm cool}/t_{\rm ff}$ and gravitational acceleration ($g$) for the runs
with NFW and NFW+BCG potentials. The plots show that the effect of BCG is felt only within the central 50 kpc of the cluster. The density and temperature in the core are nearly double when
the BCG potential is included. Bottom right panel of Figure \ref{fig:comp} shows that the inclusion of BCG potential makes the gravitational acceleration rise sharply at small radii. This
affects the $t_{\rm cool}/t_{\rm ff}$ profile in the inner regions ($r \lesssim 20$ kpc). The BCG potential, as we discuss later, also shortens the feedback response time and 
prevents min(\tr) in NFW+BCG runs from falling below  $\approx 2$.
The initial min($t_{\rm cool}/t_{\rm ff}$) is $8.8$ for the NFW run while it is $9.8$ for the NFW+BCG run.   

\begin{figure*}
 \includegraphics[width=18cm, height=13cm]{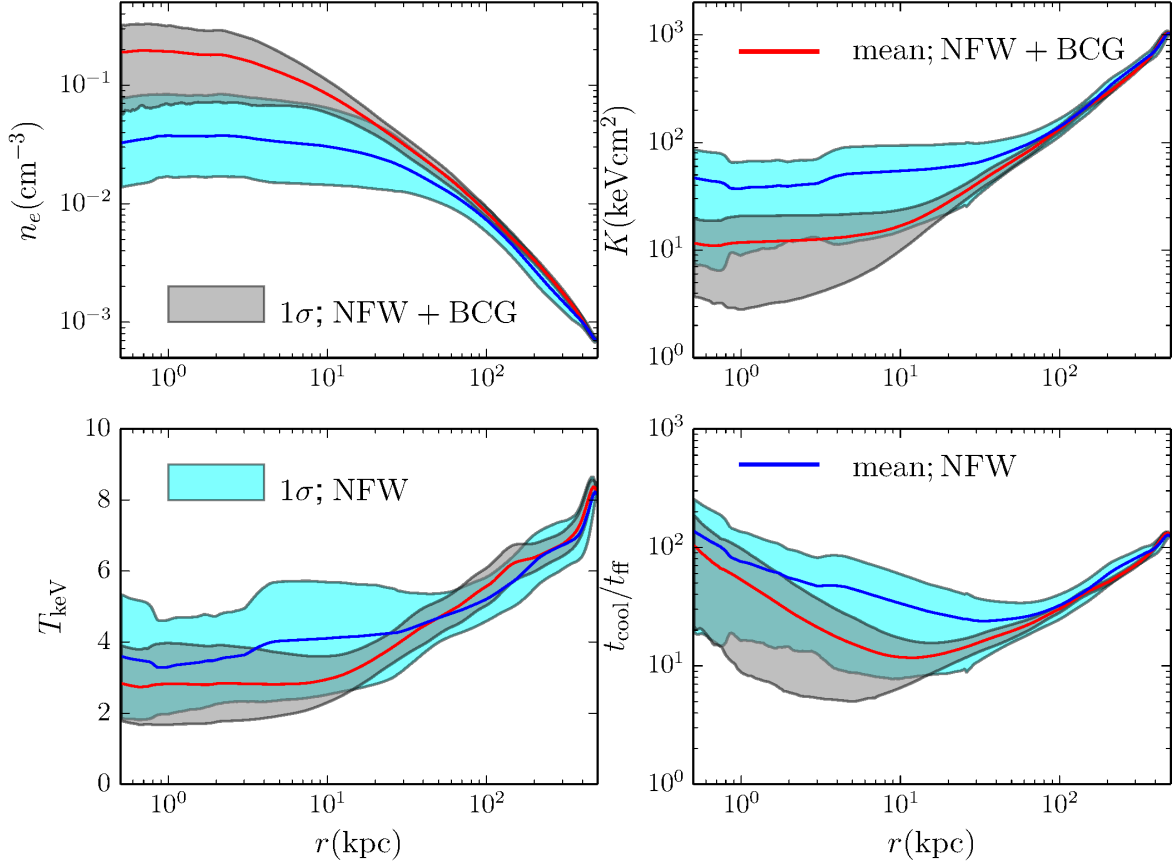}
  \caption{Angle-averaged, emissivity-weighted electron number density, temperature, entropy and $t_{\rm cool}/t_{\rm ff}$ profiles and  $1\sigma$ spread at all radii  calculated from
  the mean  value for the X-ray gas (0.5-10 keV) from
  1-4 Gyr. 
  The electron number density is on the higher side for
  the NFW+BCG run compared to the NFW run throughout  evolution  but the temperature profiles are similar for both cases.
} 
  \label{fig:all_time}
\end{figure*}

Figure \ref{fig:all_time} shows the mean and $1-\sigma$ spread in the angle-averaged, emissivity weighted electron number density, temperature, entropy and
$t_{\rm cool}/t_{\rm ff}$
profiles for the X-ray emitting gas (0.5-10 keV) from 1-4 Gyr for NFW and NFW+BCG runs. The mean and $1\sigma$
spread about the mean are calculated at each radius for different quantities between 1-4 Gyr. 
The density plot in Figure \ref{fig:all_time} shows that the average core density and the $1\sigma$ spread about the mean are higher for the
NFW+BCG run as compared to the NFW run. This higher electron number density is expected as the deeper potential well makes
it difficult for AGN jets to remove the gas from the
cluster core (due to the addition of the BCG potential). 

Unlike density, temperature  does not show any significant difference for the NFW and NFW+BCG runs. 
Although, there was a difference in the initial cluster core temperature (see Figure \ref{fig:comp}),  radiative
cooling and AGN heating cycles remove  this difference during the course of long term evolution. As a result of electron number density being
higher  for the NFW+BCG run, entropy ($K=T_{\rm keV}/n_e^{2/3}$) is  correspondingly lower. The $1\sigma$ spread of entropy about the mean has a  small overlap in the cluster core while they almost
lie on top of each other at larger radii. This shows that the effect
of BCG is only felt within the central $50$ kpc of the cluster. Outer regions are largely unaffected by the presence of BCG at the cluster
centre. 

The evolution of $t_{\rm cool}/t_{\rm ff}$ profile in Figure \ref{fig:all_time} shows a behavior similar to entropy for both
NFW and NFW+BCG runs. 
 Similar to entropy, the average $t_{\rm cool}/t_{\rm ff}$ profile
separates below 50 kpc for the NFW and NFW+BCG runs. Owing to a shallower potential well,
AGN jets are able to evacuate the core in the NFW run easily, leading to a longer cooling time. Despite having a longer free-fall
time ($t_{\rm ff}$), the longer cooling time ($t_{\rm cool}$) leads to a $t_{\rm cool}/t_{\rm ff}$ ratio for the NFW run 
which is above
that of the NFW+BCG run with  small overlap in the core. For a shallower potential well, AGN jets are able to cause overheating
out to   larger distances from the centre. 

\subsubsection{Jet Power and X-ray Luminosity}
\label{sec:pcav_lx}
\begin{figure}
 \centering
 \includegraphics[width=0.48\textwidth]{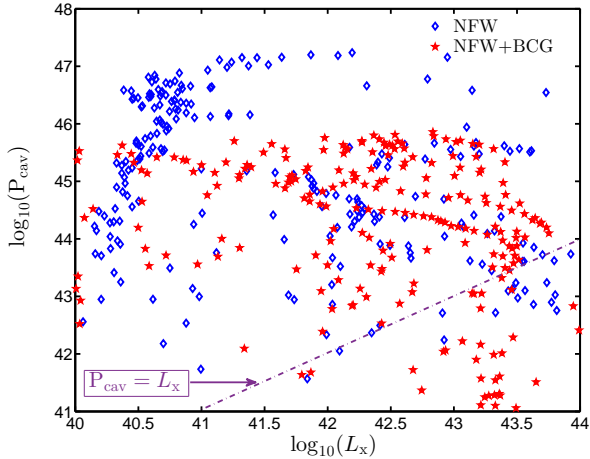}
  \caption{ Cavity/jet power (P$_{\rm cav}$) plotted against X-ray luminosity of the core ($r<30$ kpc, $L_{\rm x}$; $0.5-2$ keV) for the run
  with (stars) and without (diamonds) the BCG potential. The dot-dashed purple line shows the locus of ${\rm P_{cav}}=L_{\rm x}$. 
  With most of the points lying above the
  ${\rm P_{cav}}=L_{\rm x}$ line, the clusters are over-heated at most times for the chosen feedback efficiency ($\epsilon$). }
  \label{fig:lx_pcav}
\end{figure}

One of the ways to look at cooling and heating in clusters is to compare the core X-ray luminosity (a crude measure of cooling) and jet/cavity power 
(a crude measure of heating). 
Figure \ref{fig:lx_pcav} shows the variation of cavity power (P$_{\rm cav}$), a proxy for jet power, with the core ($r<30$ kpc) X-ray luminosity of gas between 0.5 to 2 keV
($L_{\rm x}$)
for the runs with and without the BCG potential. Cavity power is calculated as described in section 3.1.5 of \cite{dev15}. 
Most of the points lie above 
the $P_{\rm cav} = L_{\rm x}$ locus, showing that the cluster is overheated at most times for our choice of accretion efficiency in both cases. 
Although not apparent from this plot, P$_{\rm cav}$ and $L_{\rm x}$ show an anti-clockwise
cyclic behaviour in P$_{\rm cav}-L_{\rm x}$ space as anticipated by \citealt{sun09, mcd10}. 
The dense cooling cluster cores with small jet power to begin with,
 start accreting at a high
rate after the condensation and infall of cold gas. A large jump in accretion rate leads to an increase in the jet power and overheating of the core. After a while, with 
a suppressed accretion rate the core cools and stage is set for another heating cycle. Further, in the NFW+BCG run, extreme heating events are absent compared to the NFW case. This is a consequence of a deeper potential well in the NFW+BCG run. Also unlike the NFW+BCG run, the NFW run spends significant 
time with P$_{\rm cav}>10^{46}$ erg s$^{-1}$  (see also Fig. \ref{fig:jp_cg}).

Another important point that is obvious from Figure \ref{fig:lx_pcav} is that the correlation between the cavity power and 
core X-ray luminosity is rather weak, and there are hysteresis cycles (a key focus of \citealt{dev15}).
The correlation is expected to be much tighter in single phase Bondi accretion (\citealt{dev17}). Because of the
episodic nature of cold gas accretion, for the same core X-ray luminosity (or equivalently min[\tr]), very different cavity powers are seen.          

\subsubsection{Jet Power, min($t_{\rm cool}/t_{\rm ff}$) \& Cold Gas Mass}

\begin{figure*}
 \centering
 \includegraphics[width=1.0\textwidth]{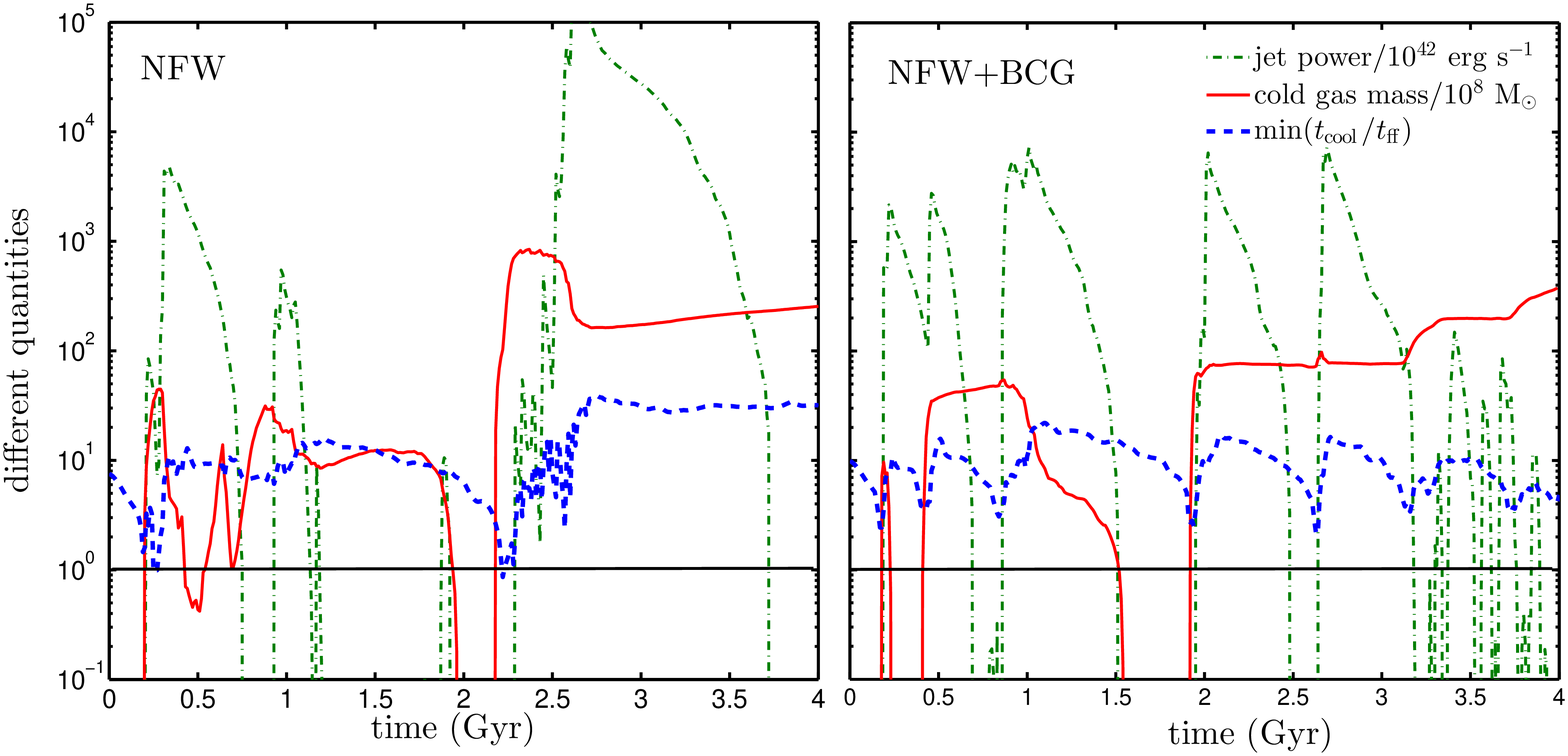}
 \caption{ Jet power (dot-dashed green line), cold gas mass (solid red line) and min($t_{\rm cool}$/$t_{\rm ff}$) (blue squares) for the  
 NFW (left panel) and the NFW+BCG (right panel) run. The time duration between consecutive jet events is
 different in the two cases with the NFW run having more powerful and longer jet events compared to the NFW+BCG run. The cold gas mass at
 the end of 4 Gyr differs in the two cases by a factor of less than 2. 
 The AGN jets are more disruptive in the NFW run in comparison to the NFW+BCG run.   
 While in the NFW run min($t_{\rm cool}/t_{\rm ff}$) goes below 1, for NFW+BCG it never drops below 2.}
\label{fig:jp_cg}
\end{figure*}

Figure \ref{fig:jp_cg} shows the time evolution of jet power, cold gas mass and min($t_{\rm cool}/t_{\rm ff}$) for  the NFW and NFW+BCG runs. 
Note that in the run with just the NFW potential (left panel), there are fewer, more powerful jet events of longer duration than
in the NFW+BCG run (right panel). As the NFW potential is shallower compared to NFW+BCG potential, the jets are able to cause
greater disruption. 
This leads to more frequent and small duration radiative cooling and AGN heating cycles for the
 NFW+BCG run as compared to the NFW run. In both cases, we see a rotationally dominant, stable cold gas torus forming in
the central few kpc, as has been reported in several works (e.g. \citealt{gas12b, li15, dev15}). By the end of 4 Gyr the amount of cold gas
exceeds $10^{10}$ M$_\odot$ in both cases, with most of the cold gas localized in the massive torus.         

The NFW run shows larger min($t_{\rm cool}/t_{\rm ff}$) values after  the jet events compared to the NFW+BCG run.
Table \ref{tab:tab1} shows that the NFW run has min($t_{\rm cool}/t_{\rm ff}$) below
10 for $47\%$ of the run time, while for the NFW+BCG run min($t_{\rm cool}/t_{\rm ff}$) is below 10 for a higher fraction ($55\%$) of the
time. The fraction of time spent with min(\tr)$<5$ is $14\%$ in both cases. Therefore, we expect only a small number of clusters with min(\tr)$<5$.
The range of  min($t_{\rm cool}/t_{\rm ff}$) for the NFW run is 1-30 while it is 2-22 for the NFW+BCG run. 
Due to stronger gravity in latter, the feedback response time is shorter; consequently, min(\tr) in NFW+BCG case never drops below 2 while in the NFW case it can drop below 1. 
For the same reason, jet power in the NFW+BCG run does not reach as high as in the NFW run; heating phase starts quickly before a lot of cooling 
(followed by large jet power) occurs. The
min(\tr) ratio and jet power are not perfectly anti-correlated.  
There are times when
accretion of gas clouds lingering from previous cycles leads to strong jet outburst even when the core is not back to  
min(\tr)$<10$. At times, multiple such
events can occur in quick succession, especially when min($t_{\rm cool}/t_{\rm ff}$) is close to 10.

\subsection{Effects of Stellar Gas Depletion}

\begin{figure}
 \centering
 \includegraphics[width=0.5\textwidth]{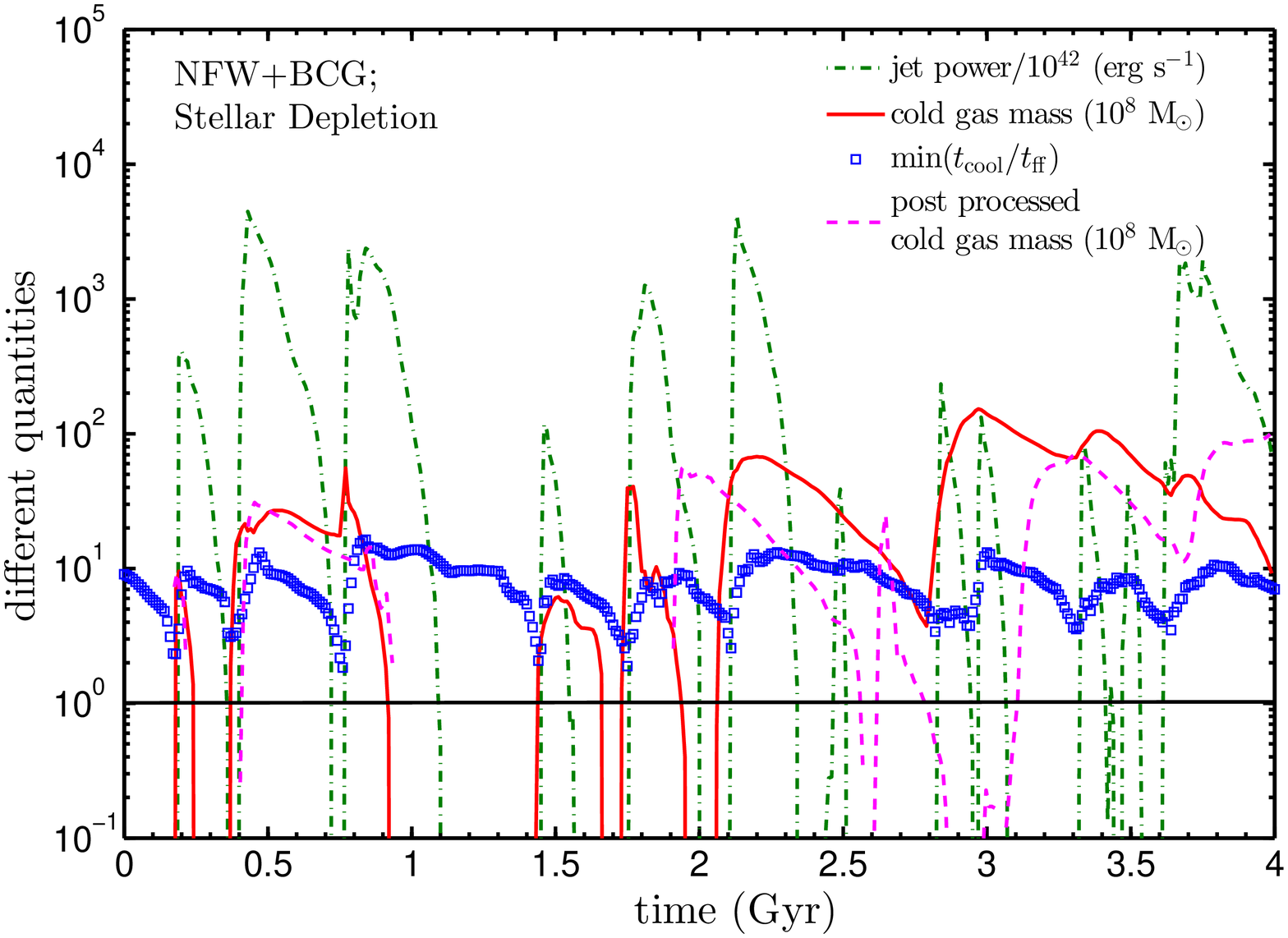}
 \caption{ Jet power (dot-dashed green line), cold gas mass (solid red line) and min($t_{\rm cool}$/$t_{\rm ff}$) (blue squares) for 
the NFW+BCG run with stellar gas depletion. The magenta dashed line shows the post processed (using Eq. 10 in \citealt{dev15}) 
cold gas mass with $\tau=200$ Myr 
for the NFW+BCG run without stellar depletion. Jet power and min($t_{\rm cool}$/$t_{\rm ff}$) show similar values to the NFW+BCG run 
without stellar depletion, but the time duration between consecutive jet events is shorter. The total cold gas 
at the end of 4 Gyr is an order of magnitude smaller in comparison to the NFW+BCG without stellar gas depletion (see the right panel of Fig. \ref{fig:jp_cg}). 
Note that the lowest min(\tr) is always above 2,  like in the NFW+BCG case without stellar gas depletion.}
\label{fig:SD}
\end{figure}

One of the  problems with recent hydrodynamic simulations of AGN feedback in galaxy clusters is the formation of a massive torus in
the central few kpc of the cluster core (\citealt{gas12b, li15, dev15}),
which is generally on the higher end of the observed cold gas mass spectrum (this of course depends on the choice of $\epsilon$). The cold 
gas in the torus is dominated by rotation and is decoupled from the feedback cycle. However, observations show that only a few clusters 
like Hydra (\citealt{ham14}) have rotating cold disks extending few kpc. In most clusters no such prominent structure is observed and most of the 
cold gas is in extended filaments (\citealt{rus16,rus17}). To ameliorate the problem of excess cold gas mass in our simulations, we  
include a simplified model for the depletion of cold gas due to star formation as described in
section \ref{sec:stell}. 

 For the NFW+BCG run with stellar gas depletion (labeled NFW+BCGd) the mass flow rate across the inner boundary is $7.1$ M$_\odot$yr$^{-1}$, comparable to that in the run without depletion
 (see Table \ref{tab:tab1}). This
 means that star formation primarily affects  the rotationally dominant cold gas component (torus) while the
 radially dominant component with a free-fall time shorter than the depletion time ($\tau$; see Eq. \ref{eq:mass_source}), 
 which controls the feedback cycle, remains largely unaffected.    

Figure \ref{fig:SD} shows the evolution of jet power, cold gas mass and min($t_{\rm cool}$/$t_{\rm ff}$) for the NFW+BCGd run. 
The dashed magenta line shows the total cold gas mass when cold gas is depleted in post processing (as in \citealt{dev17}) for
the NFW+BCG run, while the solid red line shows the total cold gas mass for the NFW+BCGd run.
In both cases, the cold gas depletion time, $\tau$, is 200 Myr. Both approaches show the dynamic nature of the amount 
of cold gas, with the peak cold gas mass at $\approx 10^{10}$ M$_\odot$. This is unlike the run without cold gas depletion in which the 
total cold gas mass gets saturated after 2 Gyr at $\gtrsim 10^{10} M_\odot$ (see the right panel of Fig. \ref{fig:jp_cg}). 
With stellar depletion, 
the total cold gas in the core lies in the range of observed cold gas mass in cool cluster cores (see Figure 8 in \citealt{dev17}).   

The evolution of jet power with time for the NFW+BCGd  run in Figure \ref{fig:SD} shows significant difference 
from the NFW+BCG run in Figure \ref{fig:jp_cg}. While the peak jet power is roughly of the same order for the two runs, the duration of 
each jet event is smaller in the run with stellar gas depletion. When cold gas gets depleted because of star formation, AGN jets encounter 
less resistance (cold clumps help energy deposition in the core by inducing local turbulence; c.f. Fig. \ref{fig:metal}) on their way out 
from the cluster core. As a result, they deposit most of their energy at larger distances. Cluster core is not overheated despite a large jet 
power, and the cluster core maintains its cool characteristics for longer. This leads to frequent jet events of shorter intervals.

The evolution of min($t_{\rm cool}/t_{\rm ff}$) ratio with time for the run with stellar gas depletion in Figure \ref{fig:SD} is as 
expected. Right before a major jet event, this ratio dips below 10 signaling a cooling phase in the cluster core. The cooling phase
is followed by a strong accretion phase which gives rise to a powerful jet outburst. This heats up the core, pushing 
min($t_{\rm cool}/t_{\rm ff}$) above 10, marking the completion of one cooling-heating cycle. This is repeated multiple 
times during our simulation. The min(\tr) ratio fluctuates between 2-20 during the course of the simulation, with 77\% of the time it lying 
below 10 but only $19\%$ of the time  below 5.  Like the NFW+BCG run with no depletion, even here the min(\tr) ratio never drops below 2.

\subsection{Turbulence in Cool-Core Clusters}

\begin{figure}
 \centering
 \includegraphics[width=0.45\textwidth]{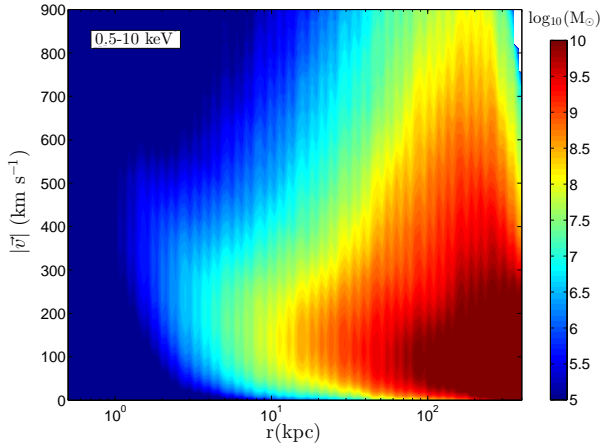}
  \caption{Time-averaged (from 1-4 Gyr) velocity-radius distribution of the X-ray gas (0.5-10 keV) for the NFW+BCG run without stellar
  depletion.
  The plot shows the $|{\bf v}|-r$ mass distribution - $d^2M/(d\ln |{\bf v}| d\ln r)$ ($\Delta v=10$ km s$^{-1}$, $\Delta \log_{10} r=0.03$
  are the bin sizes).
In the central regions ($r<60$ kpc), most of the hot gas mass (of the order of $10^{10}$ M$_\odot$) is in the
velocity range of 0-400 km s$^{-1}$. However, small amount of gas ( $\sim 10^7$ M$_\odot$) has velocity going up to 1000 km s$^{-1}$. A similar
velocity-radius map of the X-ray gas is obtained for the NFW+BCG run with stellar depletion.}
  \label{fig:xvel}
\end{figure}

\begin{figure*}
 \centering
 \includegraphics[width=3.5in,height=2.5in]{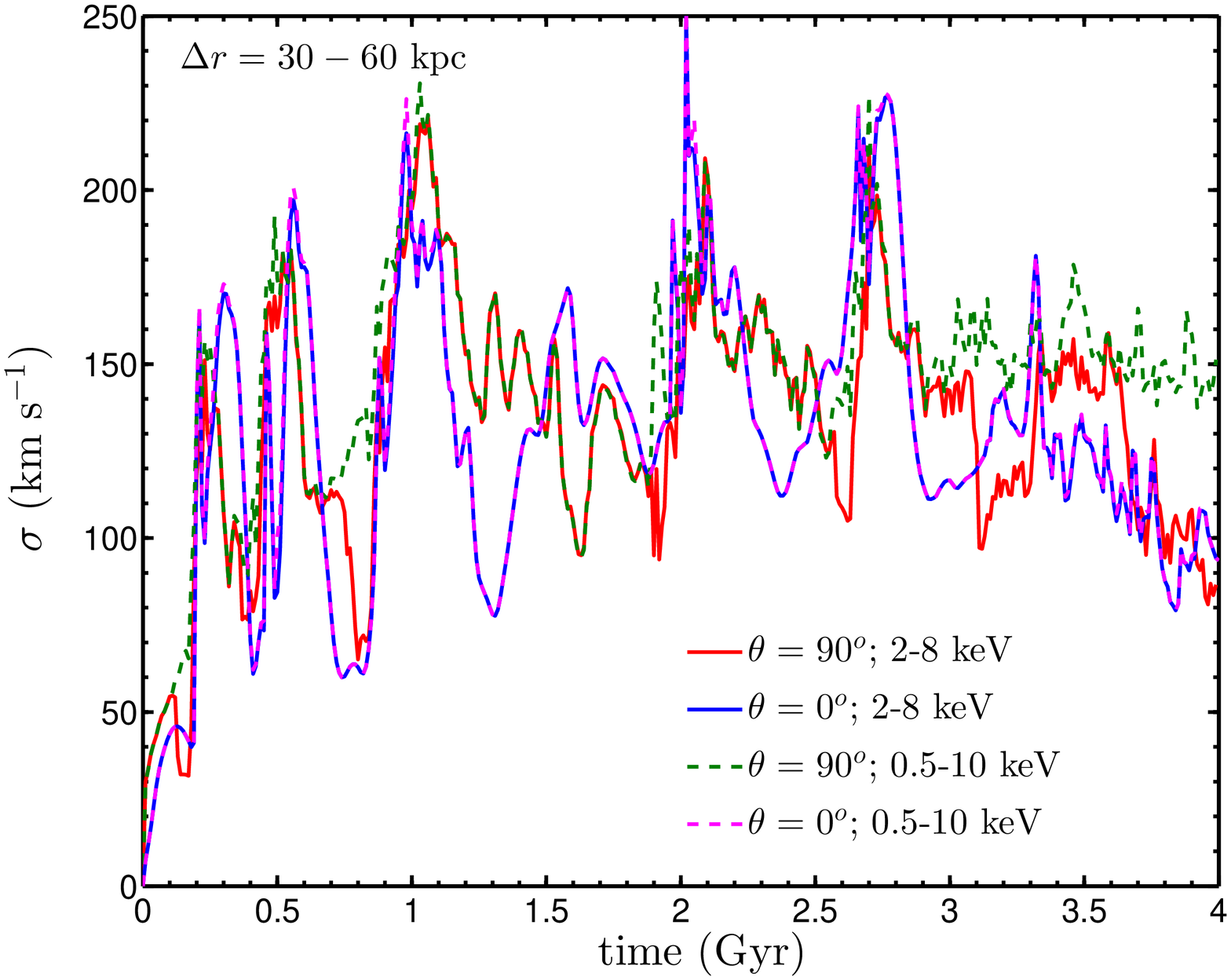}
 \includegraphics[width=3.5in,height=2.5in]{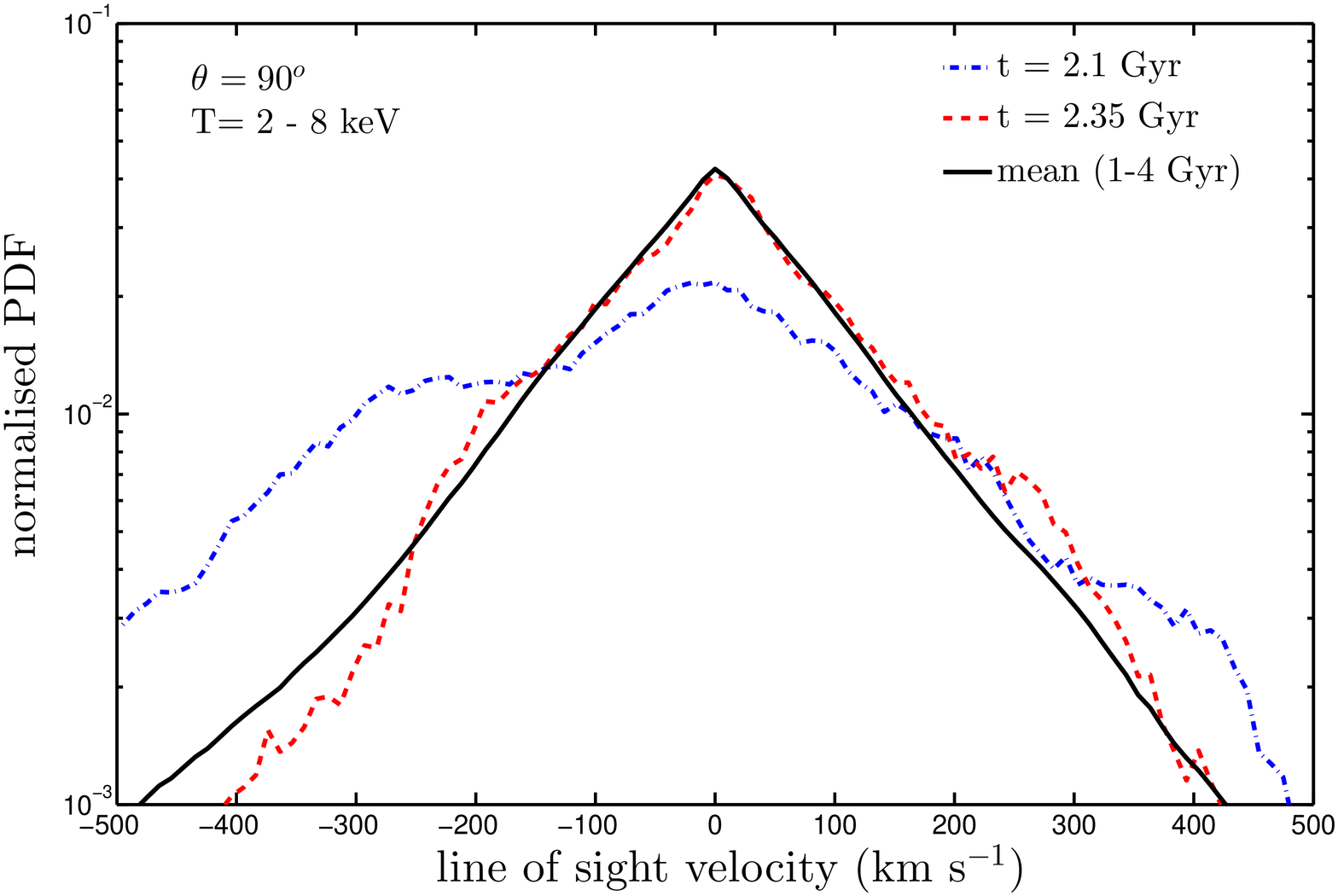}
        \caption{Left panel shows the line of sight velocity dispersion (LOSVD) with time for 2-8 keV plasma (a proxy for Fe XV line emission) 
        and total X-ray gas (0.5-10 keV) within
        30-60 kpc of the NFW+BCG run without stellar gas depletion. The solid lines (blue line is parallel to jet direction (z-axis) while red line is perpendicular to jet direction) represent
         the LOSVD for 2-8 keV  gas. Dashed lines (green line is perpendicular to the jet direction while magenta line is parallel to the jet direction) show the LOSVD
        for the total X-ray gas (0.5-10 keV). Right panel shows the probability distribution function (PDF) of the volume-weighted line of sight
        velocity perpendicular to the jet direction for the 2-8 keV gas. The solid black line is the time-averaged PDF between
        1-4 Gyr, while the dashed red and dot-dashed blue lines correspond to the PDF
        at $t=2.35$ Gyr and $t=2.1$ Gyr, respectively. These times correspond to a trough and a peak in
        the LOSVD as seen in the left panel. Note the presence (absence) of high velocity tails at 2.1 (2.35) Gyr.}
  \label{fig:disp}
\end{figure*}

AGN-ICM interaction gives rise to turbulent motion of the gas in the ICM. In our previous paper (\citealt{dev15}), we looked at the cold gas
kinematics in the cluster core due to AGN-ICM interaction. Here, we look at the motion of  X-ray emitting hot gas (0.5-10 keV) in the cluster core and
compare it with the {\it Hitomi} results for the Perseus cluster (\citealt{hit16}).

Figure \ref{fig:xvel} shows the average (from 1-4 Gyr) velocity-radius distribution ($d^2M/d\ln|{\bf v}|d\ln r$ ) of the X-ray gas
(0.5 - 10 keV) for  the NFW+BCG run. The central 100 kpc hot gas distribution shows that most of the hot gas mass  ($\gtrsim 10^{10}$ M$_\odot$)
lies in the velocity range of 0-400 km s$^{-1}$. For 100-300 kpc radial range, the hot gas velocity has a broader distribution
with the velocity range expanding to 0-500 km s$^{-1}$. There is also a small fraction of gas with velocity reaching beyond 900
km s$^{-1}$. The small fraction of gas attains this high value due to AGN jets. Beyond 300 kpc, most of the mass has velocity in the range of 0-200 km s$^{-1}$.

The left panel of Figure \ref{fig:disp} shows line of sight velocity dispersion (LOSVD) with time within 30-60 kpc  of the cluster center for the X-ray gas in our NFW+BCG run. We show the results for all X-ray gas (0.5-10 keV) 
and for 2-8 keV plasma in the same spherical shell. The latter range corresponds to the gas responsible for {\sl Fe XXV} 6.6 keV line probed by {\it Hitomi} in the core of 
Perseus cluster. 

The LOSVD is calculated for the velocity component parallel
to the y-axis (perpendicular to jet injection direction; $\theta = 90^o$) and z-axis (parallel to jet injection direction; $\theta=0^o$). Other than a slight divergence in the LOSVD of the X-ray gas at late times ($t>3$ Gyr) for $\theta=90^o$, the results for the hot and total X-ray gas, and between different  
orientations are very similar. The timing of
the sharp rise and fall in the velocity dispersion correlates with the rise in jet power (see the right panel of Fig. \ref{fig:jp_cg}). The AGN jet outbursts push the
LOSVD from 100 km s$^{-1}$ range to $\sim 250$ km s$^{-1}$. As the jet activity dies down, velocity dispersion drops back
to the 100 km s$^{-1}$ level. 
The LOSVD 
is in the same range as earlier independent simulations (\citealt{li16,lau17}).  The $\theta=90^o$ LOSVD is higher in the quiescent state for 0.5-10 keV 
compared to 2-8 keV because of a large fraction of infalling/rotating soft X-ray emitting gas ($<$ 2 keV) in the mid-plane at these epochs.

The right panel of Figure \ref{fig:disp} shows the probability distribution function (PDF) of the line of sight velocity (LOSV)
perpendicular to the jet direction ($\theta=90^o$)
for the hot X-ray gas (2-8 keV) within 30-60 kpc (to compare with Hitomi observations of the Perseus core).
The solid black line shows the time-averaged PDF (averaged from 1-4 Gyr), and dashed red and dot-dashed blue lines show the PDF
at 2.35 Gyr and  2.1 Gyr.These times correspond to a trough and  a peak, respectively, in the LOSVD plot in the left panel. The PDF for
the LOSV peaks close to 0 km s$^{-1}$, showing the absence of significant bulk flows in the hot phase.
During strong AGN jet activity, the PDF of the LOSV has an extended high velocity ($\gtrsim$ 300 km s$^{-1}$) tail which is absent otherwise.
LOSV distribution is similar along the jet direction ($\theta=0^o$), except with somewhat higher velocity tails when the jet is active.

\subsection{Spread Of Metals  By AGN Jets}

\begin{figure}
 \centering
 \includegraphics[width=0.50\textwidth]{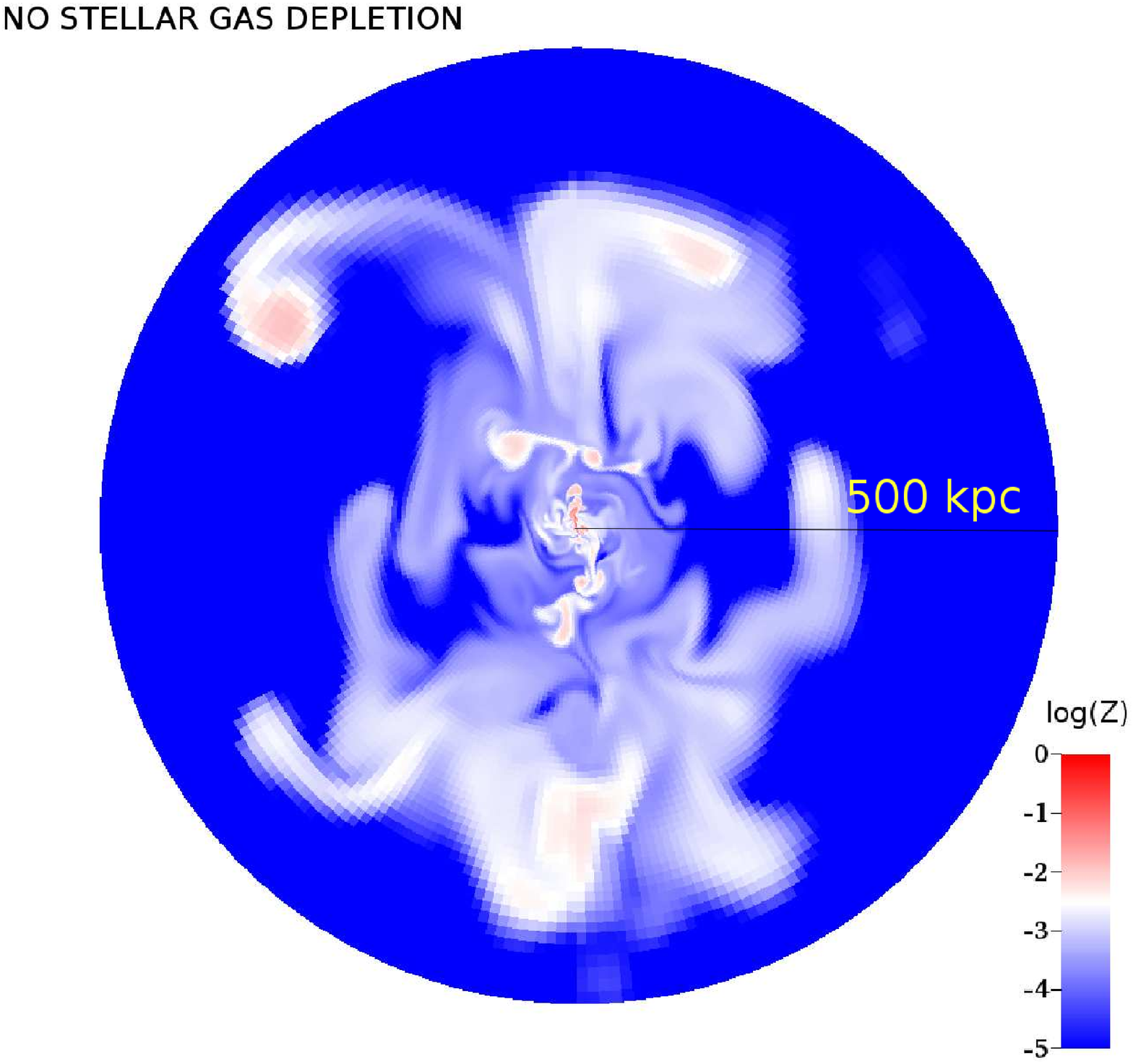}
 \includegraphics[width=0.50\textwidth]{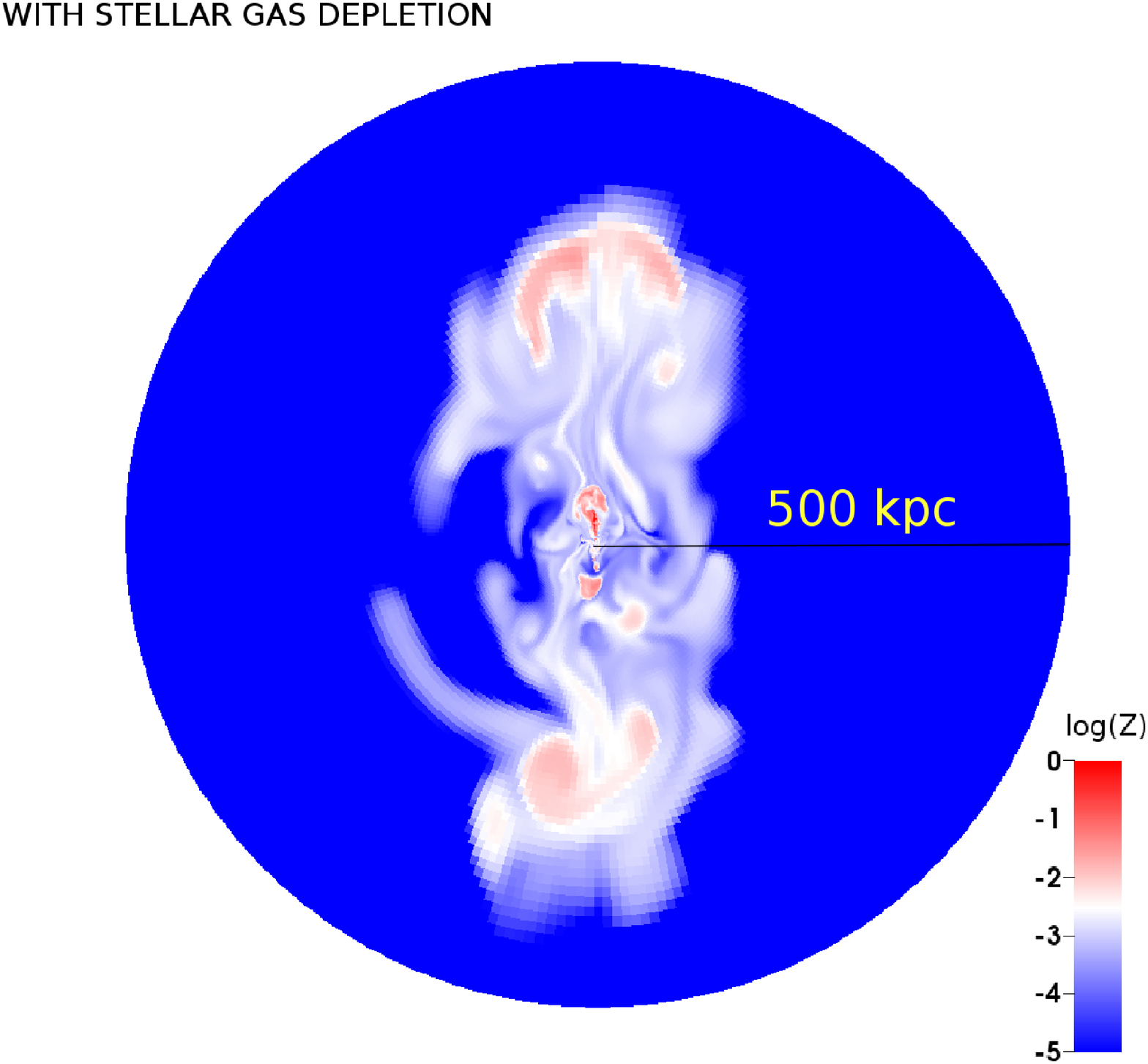}
  \caption{Metallicity at 4 Gyr for the NFW+BCG runs with (bottom panel) and without (top panel) stellar gas depletion. 
  Note that the lateral mixing of metals is more extensive in absence of gas depletion.
  }
  \label{fig:metal}
\end{figure}
\begin{figure}
 \centering
 \includegraphics[width=0.5\textwidth]{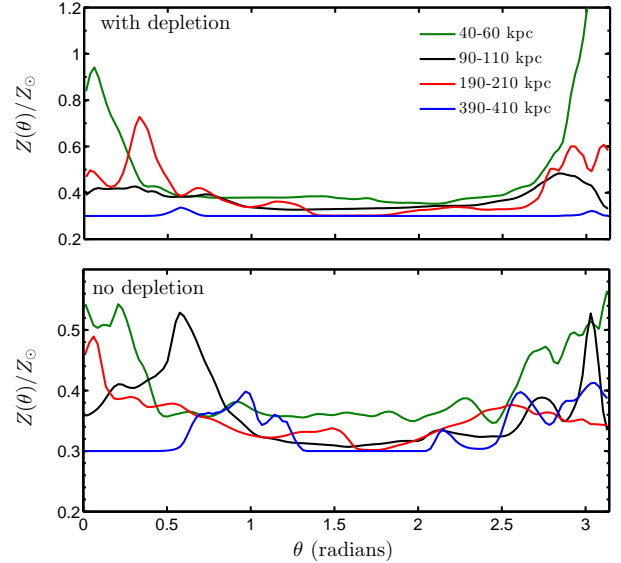}
  \caption{Angular spread of  metals due to AGN jets at 4 Gyr (see Eq. \ref{eq:ang_dist}) for shells at different
radii ($\Delta r = 20$ kpc) for the NFW+BCG runs without (lower panel) and with (upper panel) stellar depletion.
The peaks at $\theta=0$ and $\pi$ show that jets are unable to disseminate metals in the equatorial
regions to large radii. The metallicity peaks in the polar regions are higher at smaller radii in the run with stellar 
gas depletion as compared to the run without depletion. A metallicity floor of $0.3 Z_\odot$
is added  to these profiles.}
  \label{fig:ang}
\end{figure}
\begin{figure}
 \centering
 \includegraphics[width=0.5\textwidth]{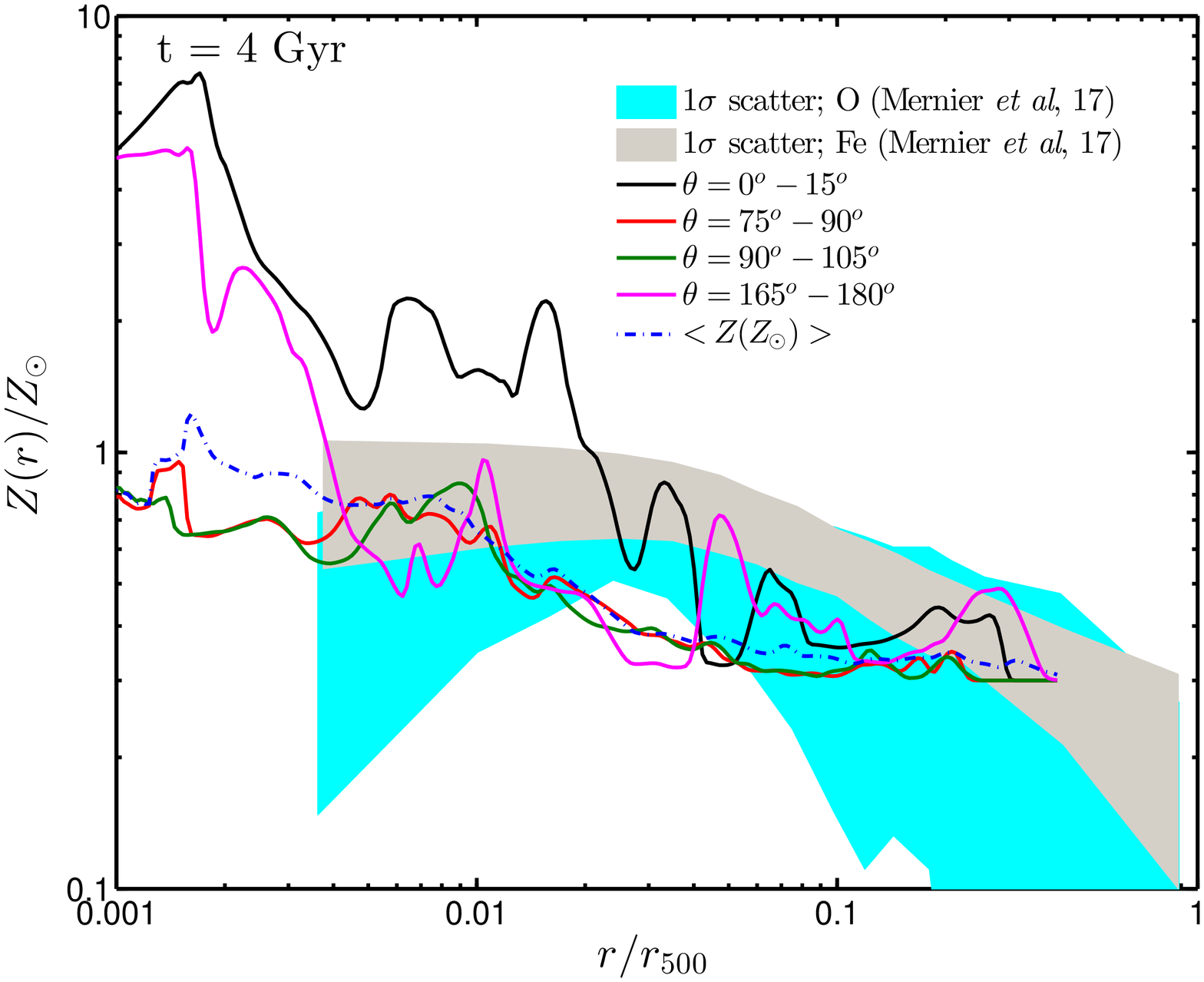}
  \caption{Angle-averaged metallicity profile (dot-dashed blue line) at 4 Gyr for the NFW+BCG run without stellar gas depletion. In the core ($r < 0.02r_{500}$) metallicity rises sharply above the background metallicity of $0.3Z_\odot$ while beyond $0.02r_{500}$ AGN jets seems to have no impact on the metallicity. The grey shaded region is the 1$\sigma$ spread about the mean for Fe and the cyan shaded region is 1$\sigma$ spread about the mean for O as a function of radius for the full  sample (cluster+groups) in \cite{mer17}. Observations show that the metallicity is elevated for much larger radii ($r < 0.1r_{500}$) in clusters in comparison to our simulation. Solid lines show the metallicity profiles within $15^o$ parallel and perpendicular to the jet direction. The average metallicity profile closely follows the equatorial profiles because of a larger
  solid angle. A metallicity floor of $0.3 Z_\odot$ is added to these profiles.}
  \label{fig:mrm}
\end{figure}

AGN jets help transport the metals from star forming core regions (and also metals produced by type Ia supernovae in the BCG) in cool core clusters. Here we use a simplified model
of metal transport in which all metals are injected close to the center in the jet injection region. Our model should be fine to quantify metal transport in cool-core clusters due to AGN jets. 

Figure \ref{fig:metal} shows the
extent of metal distribution due to AGN jets in the NFW+BCG  simulations, at the end of 4 Gyr for the runs with (left panel) and without (right panel) stellar depletion.
These plots show that AGN jets can spread metals  beyond 400 kpc in both cases, predominantly in the jet direction. 
There is a noticeable difference between metallicity distribution at larger radii in the runs with and without gas depletion. Jets encounter less  resistance 
due to clumpy cold clouds in the simulation with cold gas depletion and therefore travel mostly along the direction of jet injection without much dispersion in the transverse directions. Without stellar depletion the metal distribution is much more laterally extended at large radii because of vorticity generation at the hot-cold interface of the cold gas clouds (see the upper panel of Figure \ref{fig:metal}). 

Figure \ref{fig:ang} shows the angular distribution of mass weighted metallicity
(Eq. \ref{eq:ang_dist}) of the ICM for different radial shells ($\Delta r = 20$ kpc) at 4 Gyr for the NFW+BCG runs with and without stellar gas depletion. The distribution shows bimodality
with metallicity peaking near the polar regions, as expected from Figure \ref{fig:metal}. However, the metallicity peak is higher by a factor of 2 for smaller radii with 
stellar depletion as compared to the run without stellar depletion, again reflecting higher lateral mixing in the latter case.  

Figure \ref{fig:mrm} shows the angle-averaged distribution of metals as a function of radius at 4 Gyr (Eq. \ref{eq:metal_radius}) for the NFW+BCG runs 
without stellar gas depletion. Metallicity shows a steep rise in the cluster core ($r<0.02r_{500}$), while beyond $0.02r_{500}$ there is only $\lesssim 2\times$ change in 
metallicity over the background 0.3Z$_\odot$. This shows that the impact of AGN jets in distributing the metals is limited to the cluster core. The 
shaded regions depict the $1\sigma$ scatter of metallicity about the mean (cyan is for O, grey is for Fe) as a function of radius for the sample of clusters and 
groups in \citealt{mer17}. The observations show a more extended metal distribution (till $0.1r_{500}$) as compared to our simulations.  

\section{Discussion}
\label{sec:dis}

Recent observations of galaxy clusters have thrown up many challenges for
the simulations of cold mode feedback in cool cores.
Two of the key challenges are: (i) the near absence of cores with min(\tr)$<$10 (\citealt{hog17, hog17a}; see however, \citealt{voi15,lak16,pul18}), unlike
smaller ratios (down to unity) seen in simulations (albeit for a short time); and (ii) the absence of massive rotating cold tori in observations, which are routinely seen within the central few kpc
of cool core simulations, except in Hydra A (\citealt{ham14}). In light of these discrepancies, we have incorporated two effects in our 
simulations to see if simulations can be reconciled with observations: (i) a central BCG potential; and (ii) a simple model for gas depletion due to 
star formation. We study the effects of these new physical ingredients on the long term evolution of cluster cores. Additionally, we quantify other important X-ray observables
such as metallicity and level of turbulence in the X-ray emitting gas in cool cluster cores.

\subsection{How much below 10 does min($t_{\rm cool}/t_{\rm ff}$) fall?} 
Pinning the gravitational acceleration to that of the central BCG at small radii, 
\citealt{hog17, hog17a} argue that 
almost none of the cool  cluster cores go below the  
min($t_{\rm cool}/t_{\rm ff}$) $=10$ threshold for the presence of cold gas motivated by simulations. While \citet{hog17a} use a sample of 33 H$\alpha$ line emitting galaxy clusters and find one core with min(\tr) slightly below 10 (they quote a range in min[\tr] of 10-35),
more recent observations of 23 cool cores with confirmed detections of CO-emitting gas (\citealt{pul18}) find 5 systems below 10 (although still above 7). In this sample 10 out of 23 CO-emitting cool cores have min$(t_{\rm cool}/t_{\rm ff})$ between 8-12 and one system below 8. Further, the latest observations of the 40 low mass halos (galaxies) by \citet{bab18} show that min($t_{\rm cool}/t_{\rm ff}$) falls to as low as 5. In this sample 8 out of 40 systems have min$(t_{\rm cool}/t_{\rm ff}) \leq 10$. \citet{voi15}, using a singular isothermal sphere model (with a fixed velocity dispersion of 250 km s$^{-1}$) 
for the BCG potential, found a min(\tr) ratio in the range 5-20 for cool core clusters. Following the method of \citet{voi15}, \citet{ewa17} find that out of five galaxy groups with jets, four have min(\tr) $<15$ with lowest value at 7.4. Thus, the disagreement between cool-core observations and simulations highlighted in \citet{hog17a} does not appear to be serious. 

\begin{figure}
 \centering
 \includegraphics[width=0.49\textwidth]{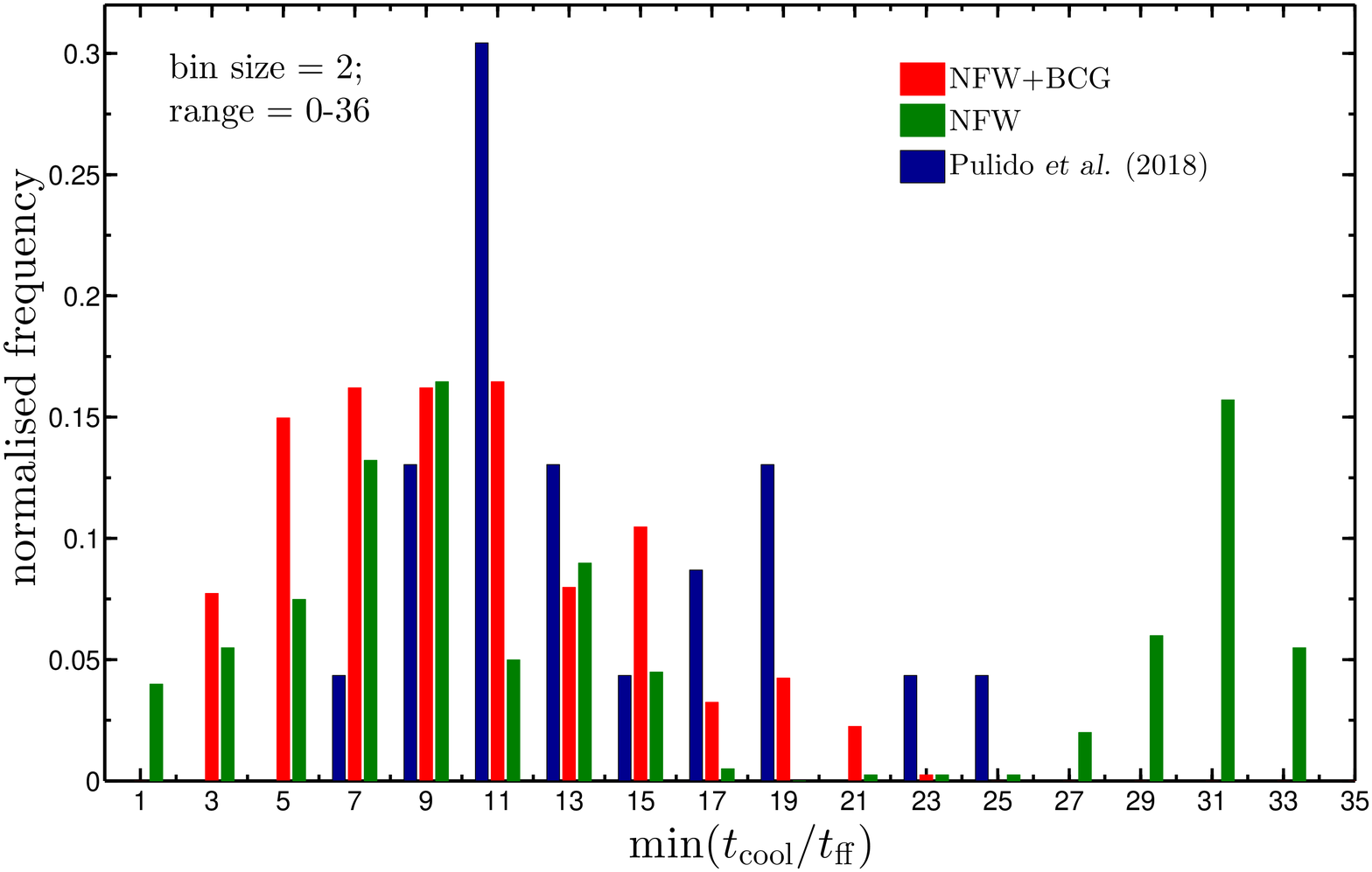}
  \caption{The normalized histograms for the distribution of simulated clusters (see Table \ref{tab:tab1}) and observed cluster cores that show CO emission (\citealt{pul18}) 
  as a function of min(\tr). Notice that the simulated NFW distribution is
  bimodal and min(\tr) goes to as low as unity in this case. All simulated clusters are biased toward lower min(\tr) compared to observations for our choice of $\epsilon$.}
  \label{fig:hist}
\end{figure}

The one noticeable discrepancy is the presence of snapshots in which min(\tr) falls as low as 1 for the NFW run (see the left panel of Fig. \ref{fig:jp_cg}). However, the BCG potential
somewhat alleviates this problem by shortening the feedback response time (as $t_{\rm ff}$  in the core is shorter) and preventing the core from cooling below min(\tr)$=2$. In this case 
feedback acts fast and feedback jet heating cycle starts before the core can cool too much. Moreover, for the same reason, the jet power and min(\tr) after the jet event do not increase to large values. 

Figure \ref{fig:hist} shows the histograms of min(\tr) for our NFW and NFW+BCG runs and from the cool 
core sample of \citet{pul18} of the clusters with CO detection. Clearly the NFW simulation shows occurrence of clusters with min(\tr) as low as unity. Moreover, the NFW core distribution is 
bimodal with another peak in min(\tr) occurring at $\approx 30$. Similar results are obtained for 3-D NFW cluster simulations in \citet{dev15}, which uses a smaller value of feedback efficiency 
($\epsilon$). With the inclusion of BCG there are no cores with min(\tr)$<2$ and there is no second peak in min(\tr) distribution. While the latter inference depends on $\epsilon$, the former
is a generic feature of a deeper potential well.
Thus the inclusion of a BCG potential brings the simulations in a closer agreement with cool core observations but even then the observed distribution is shifted toward higher min(\tr). The peak 
of the observed and simulated BCG distributions is within a factor of two. The disagreement at low values of min(\tr) may be due to several factors: observational biases due to sample selection
(e.g., even different samples by the same group show different distributions of min[\tr]; e.g., compare \citealt{hog17a,pul18,bab18}); low spatial resolution (see Fig. 8 in \citealt{hog17a}) and breakdown 
of spherical symmetry in the core where min(\tr) occurs; simulations do not include important physical effects such as thermal conduction and stellar feedback that can increase min(\tr).

Now we discuss the issue of the occurrence of cold gas even when min(\tr)$>10$.
Early idealized thermal instability models with heating balancing average cooling in radial shells (\citealt{mcc12,sha12,cho16}) for the hydrostatic ICM confined by NFW gravity
showed min(\tr) $\lesssim 10$ to be a necessary condition for cold gas condensation. Amount and extent of cold gas condensation is higher for a smaller min(\tr). In these models
without angular momentum, extra gas from the core drops out leaving behind a core with min(\tr)$\gtrsim 10$ (the exact value of this threshold depends on the 
\tr~profile, as investigated in \citealt{cho16}) that is no longer susceptible to multiphase condensation. 

The more realistic feedback AGN jet-ICM simulations show irregular 
cooling and heating cycles of the core, with min(\tr) falling below 10 in the cooling phase  
and rising above the threshold value after the cold-gas-driven
AGN jet overheats the core to min(\tr)$\gtrsim 10$. The fraction of time the core spends with min(\tr)$>10$ depends on the halo mass and the feedback efficiency. For a higher 
feedback efficiency and a shallower gravitational potential, the core spends a longer time in an overheated state with min(\tr)$>10$. E.g., Table \ref{tab:tab1} shows that the NFW run
with a shallower potential spends a shorter time with min(\tr)$\lesssim 10$ as compared to the NFW+BCG runs, for the same feedback efficiency. Unlike in idealized thermal instability 
simulations (and 2-D axisymmetric jet-ICM simulations), angular momentum of cold gas plays a crucial role  
in 3-D simulations. The core can {\em retain substantial amount of cold gas even in the heating phase} as the cold gas supported by angular momentum is not quickly dispersed or converted into stars. Thus, in
realistic 3-D jet-ICM simulations the correlation between various cool-core diagnostics is not strong (e.g., see Fig. 10 in \citealt{li15} and Fig. 14 in \citealt{dev15}).
For example, Figures \ref{fig:jp_cg} \& \ref{fig:SD} show that substantial cold gas (and consequently star formation) is present even if min(\tr) is in the range 10-30 and not necessarily $<10$, 
consistent with the observations.

\subsection{Turbulence in Cool Cores}

Indirect constraints on turbulent velocities -- based on surface brightness fluctuations (e.g., \citealt{zhu14}), scattering of resonant lines (e.g., \citealt{wer09}),
comparison of optical and X-ray derived gravitational accelerations (e.g., \citealt{chu08}) -- in cool cluster cores show that the turbulent energy is $\lesssim 10\%$
of the thermal energy. X-ray line spectra from RGS (Reflection Grating Spectrometer) on {\sl XMM-NEWTON}, because of its insufficient spectral resolution, could
only put weak upper limits on non-thermal velocities (\citealt{san11}). The situation improved after the soft X-ray spectrometer (SXS)  on \textit{Hitomi}, with its superior
spectral resolution, directly measured the line of sight turbulent velocity dispersion $\approx 164\pm10~{\rm km~s}^{-1}$ within 30-60 kpc of the Perseus core (\citealt{hit16}).
The turbulent pressure is only $\sim 4$ \% of the thermal pressure, despite a fairly large jet/cavity power $\sim 10^{45}$ erg s$^{-1}$ (\citealt{bir04}; see Fig. \ref{fig:jp_cg}
for comparison).

Equating turbulent heating rate density ($\rho u_L^3/2L$; $u_L$ is the velocity measured at scale $L$) and radiative cooling rate density
($n_e n_i \Lambda$; $n_{e/i}$ is electron/ion number density and $\Lambda$ is the cooling function), we obtain that
a turbulent velocity of
\be
\label{eq:reqd_uL}
u_L \approx 450~{\rm km~s}^{-1} \left  ( L_ {\rm30} \Lambda_{-23} n_{e, 0.05}  \right )^{1/3}
\ee
is required for turbulent heating to balance radiative cooling losses, where $L_{30}$ is the driving length scaled to 30 kpc, $\Lambda_{-23}$ is
the cooling function scaled to $10^{-23}$ erg cm$^3$ s$^{-1}$ and $n_e$ is electron number density scaled to 0.05 cm$^{-3}$. This is much
larger than the 3-D velocity dispersion measured in the core of Perseus $\approx \sqrt{3} \times 164 = 285$ km s$^{-1}$. Moreover, for
cold gas condensation not to be suppressed by turbulent mixing, condensation should occur at scales larger than the driving scale
(see section 4.1 in \citealt{ban14}). While {\sl Hitomi} observations rule out turbulent heating with a driving scale $\gtrsim 10$ kpc
as the dominant heating mechanism in the core, turbulent mixing of the core and the hot outskirts and/or AGN bubble is still possible (e.g., \citealt{ban14,hil16,yan16a}).

We have quantified turbulent velocities in the hot gas in our simulations that broadly agree with observations. Figure \ref{fig:xvel} shows the velocity-radius distribution of
the hot gas (0.5-10 keV) mass and shows that most of the X-ray emitting gas has 3-D turbulent velocity $\lesssim 500$ km s$^{-1}$. Figure \ref{fig:disp} shows the 1-D
line of sight velocity dispersion in the core as a function of time (left panel) and the pdf of the LOS velocity (right panel). The turbulent velocity increases with a rise in jet
power, but only up to $\lesssim 200$ km s$^{-1}$. The velocity dispersion in the direction of the jet is slightly higher than in the perpendicular direction and the turbulent velocity
even in the quiescent state is $\gtrsim 80$ km s$^{-1}$. A weak turbulent velocity motivates other models such as intermittent shocks (e.g., \citealt{li17}) and turbulent
mixing (e.g., \citealt{ban14,hil17}) as the agents responsible for heating of the cool core.

\subsection{Metal Distribution in Cool Cores}
Observations show that the outskirts ($r> r_{2500}$, the radius within which the mean matter density is 2500 times the critical density of the universe;
for our choice of parameters, $r_{2500}=0.25 r_{200}=459$ kpc) 
of  
galaxy clusters have roughly isotropic distribution of metals 
(\citealt{tam04,fuj08, sim11, wer13}), which is close to 0.3 times the solar metallicity across different systems.
Moreover, the cool-core clusters have a rising metallicity toward the center (\citealt{deg01,lec08,ett15}). 
While AGN jets 
play a key role in metal transport in the central regions of cool core clusters, our simulations show that 
they cannot be responsible for the isotropic distribution of metals in cluster outskirts (see Fig. \ref{fig:metal}). 
Metal enrichment during the galaxy  assembly stage at redshifts $\gtrsim 1$
and mixing driven by mergers seem responsible for such an isotropic and universal metal distribution in the cluster outskirts. The observed large-scale metal enrichment 
of galaxy clusters sheds light on the cluster formation environment at higher redshifts.

Further, a quantitative comparison of our metal distribution with the observed 
metallicity shows that our metal distribution due to AGN jets is too narrowly distributed toward the center (see Fig. \ref{fig:mrm}).  This finding is similar to \citet{kan17},
who compare metal transport with and without thermal conduction in cosmological simulations of galaxy clusters including AGN feedback. In absence of thermal conduction, 
like us, they find a very
centrally peaked metallicity distribution. With thermal conduction mixing is more efficient, and metals and heat are spread out more uniformly and to larger radii (see Fig. 3 
in \citealt{kan17} and Fig. 1 in \citealt{sha09b}). Thus the shallow metallicity profiles of cool core clusters compared to our simulations (see Fig. \ref{fig:mrm})
 point to the importance of (anisotropic) thermal conduction in heat and metal transport in cluster cores.   An additional caveat is that our idealized simulations do not include 
cosmological halo mergers that can further stir the ICM, especially at higher redshifts (e.g., see \citealt{vog18}).

\section{Conclusions}
\label{sec:con}
We study  the effects of the gravity of the brightest central galaxy (BCG) and depletion of cold gas due to star formation in our simulations
to compare with the observations of cool core clusters. 
We also study the nature of turbulence in cool cluster cores and the metal distribution due to AGN jets. Following are our key conclusions:\\

\begin{enumerate}

\item The presence of BCG potential does not have an impact on the temperature in the cluster core. However, for a fixed feedback efficiency the
presence of the BCG increases the average density of hot gas in the core. A larger core density decreases the core entropy and the $t_{\rm cool}/t_{\rm ff}$ ratio,
on average (see Fig. \ref{fig:all_time}). 
 AGN jets cause greater disruption in the  core of the shallower NFW potential as compared to the NFW+BCG potential. A stronger gravity due to the central BCG makes the feedback jets 
 respond faster and prevents min(\tr) from falling below 2, discernibly higher than the minimum value with only the NFW potential  ($\approx 1$). For the same reason 
 the jet power with the inclusion of the BCG potential does not rise beyond $10^{46}$ erg s$^{-1}$.
 The min(\tr) distribution of our BCG 
 simulations is still biased toward smaller values compared to the observations (Fig. \ref{fig:hist}), but the discrepancy is at less than a factor of two level. Moreover, 3-D jet simulations produce
 cold gas with angular momentum which can exist even with min(\tr) as high as 30, in agreement with observations. Given a dispersion in the observational results and the low angular 
 resolution in the core, the discrepancy between observations and realistic 3-D feedback jet simulations is not glaring. 
    
\item Star formation, modeled with a gas depletion time of 0.2 Gyr, 
removes the cold gas present in the clumps and torus in the cluster core. This brings down the cold gas mass within the observed range (see Fig. \ref{fig:SD}). Moreover,
the outgoing
AGN jets encounter less resistance with the depletion of cold gas,  and the  transfer of heat from AGN jets to the entire cluster core is less efficient (see Fig. \ref{fig:metal}). 
This results in more frequent AGN jet events with stellar depletion.

\item The line of sight velocity dispersion of X-ray gas in the cluster core shows that the turbulence due to AGN jets is not strong enough for turbulent heating to balance
radiative cooling in cluster cores. We find the 1-D velocity dispersion to be in in the range of 80-250 km s$^{-1}$, consistent with recent observations of the
Perseus cluster by \textit{Hitomi} (see Fig. \ref{fig:xvel}). The turbulent velocity is larger when AGN jet is active, and the line of sight velocity dispersion is slightly larger along the 
jet direction rather than perpendicular to this direction during the active jet phase (see Fig. \ref{fig:disp}).

\item Gas depletion due to star formation also modulates the anisotropic metal distribution in galaxy clusters due to AGN jets as outflowing metal-rich gas faces less
hindrance from the cold gas clouds. Metals are able to travel unhindered for the most part to outer radii and so the metal
distribution is mostly confined in the jet direction. Moreover, the observed metal distribution in our simulations is too sharply peaked toward the center as compared
to the observations of cool cores (see Fig. \ref{fig:mrm}). Thermal conduction (both isotropic and anisotropic) can help spread heat and metals more uniformly and 
farther out by overcoming strong entropy stratification. This may bring the metallicity distribution in line with the observations.

\end{enumerate}

\acknowledgments
This work is partly supported by India-Israel joint research grant (6-10/2014[IC]). 
DP is supported by a CSIR grant (09/079[2599]/2013-EMR-I) and IISc RA fellowship. AB acknowledges funding from NSERC Canada through the Discovery Grant program, 
the Institut Lagrage de Paris, and Pauli centre for Theoretical Studies ETH UZH. He also thanks the Institut
d’Astrophysique de Paris (IAP), at the Institute for Computational Sciences and University of Zurich for hosting him.
We acknowledge the support of the Supercomputing Education and Research Centre (SERC) at IISc for facilitating
our use of Cray XC40-SahasraT cluster on which our simulations were carried out. DP wants to thank Naveen Yadav for useful suggestions.

\bibliography{pap4}
\end{document}